\documentstyle[12pt]{article}

\begin{document}

\begin{center}
{\Large{\bf Nonlinear spin dynamics in ferromagnets with electron-nuclear
coupling} \\ [5mm]
V.I. Yukalov$^{1,2}$, M.G. Cottam$^1$, and M.R. Singh$^1$} \\ [3mm]
{\it $^1$Center for Chemical Physics and Department of Physics and Astronomy \\ 
University of Western Ontario, London, Ontario N6A 3K7, Canada \\
$^2$Bogolubov Laboratory of Theoretical Physics \\
Joint Institute for Nuclear Research, Dubna 141980, Russia}

\end{center}

\begin{abstract}

Nonlinear spin motion in ferromagnets is considered with
nonlinearity due to three factors: (i) the sample is prepared in a strongly
nonequilibrium state, so that evolution equations cannot be linearized as 
would be admissible for spin motion not too far from equilibrium, (ii)
the system considered consists of interacting electron and nuclear spins
coupled with each other via hyperfine forces, and (iii) the sample is
inserted into a coil of a resonant electric circuit producing a resonator
feedback field. Due to these nonlinearities, coherent motion of spins can
develop, resulting in their ultrafast relaxation. A complete analysis of
mechanisms triggering such a coherent motion is presented. This type of
ultrafast coherent relaxation can be used for studying intrinsic
properties of magnetic materials.

\end{abstract}

\vspace{2cm}


\newpage

\section{Introduction}

There are several different types of spin dynamics in condensed matter,
which can be distinguished according to whether the sample studied is in
equilibrium, weak nonequilibrium, or strong nonequilibrium. Microscopic
spin oscillations in equilibrium magnetic materials are related to magnons
and are studied by scattering techniques, such as neutron [1] or
light [2] scattering. Small deviations from equilibrium, caused by an
alternating external field, are characteristic of resonance experiments,
like electron spin resonance [3] or nuclear magnetic resonance [4]. However, 
when the initial state of a spin system is made strongly nonequilibrium, 
several types of spin relaxation can occur. If there are no transverse
external fields acting on the spins, they relax to an equilibrium state by an
exponential law with a longitudinal relaxation time T$_1$. When the
motion of spins is triggered, at the initial time, by a transverse magnetic
field, the relaxation is again exponential, but with a transverse relaxation
time T$_2$ which is usually much shorter than T$_1$.

A rather different relaxation regime from a strongly nonequilibrium initial
state arises if the spin system is coupled to a resonator. This can be done
by inserting the sample into a coil connected with a resonance electric
circuit. Because of the action of the resonator feedback field, the motion of
spins can become highly coherent resulting in their ultrafast relaxation
during a characteristic time much shorter than T$_2$ [5]. This latter type of coherent spin relaxation from a strongly nonequilibrium
state in the presence of coupling with a resonator is the most difficult to
realize experimentally and to describe theoretically. Experimental
difficulties have been overcome in a series of observations of this
phenomenon for a system of nuclear spins in a paramagnetic matrix [6-10]. A
theory of the coherent spin relaxation could be based on the
phenomenological Bloch equations, but solely for the case when the process
is triggered by a sufficiently strong coherent pulse thrust on spins at
the initial time, so that spin interactions are of no importance and only the
resonator field plays a role. However, the most interesting case is when the
coherent relaxation develops in a self-organized way from an initially
incoherent state, with no external coherent pulses triggering the process.
For such a {\it self-organized coherent relaxation} spin interactions are of
crucial importance. Then the Bloch equations become inapplicable and one has
to resort to microscopic models.

A microscopic approach for describing coherent processes in spin systems has
been recently developed [11, 12] and applied to a system of
nuclear spins interacting through dipole forces. It was shown that the
main role in initiating self--organized coherent relaxation is played by
the anisotropic (so--called nonsecular) part of the dipole interactions.

In the present paper we extend the microscopic theory of coherent spin 
relaxation [11, 12] to a much wider class of materials. We consider a rather 
general Hamiltonian including both nuclear as well as electron subsystems
interacting with each other through hyperfine forces. The electrons can possess
a long--range magnetic order as in ferromagnets or ferrimagnets, and
magnetocrystalline anisotropy is taken into account. A general investigation 
of strongly nonequilibrium nonlinear processes in realistic magnetic 
materials is of interest by itself and can also be useful for many 
applications. For instance, the self-organized coherent relaxation, being 
quite different from other types of relaxations, may give additional 
information on intrinsic properties of magnetic materials. The ultrafast 
relaxation can be employed for repolarizing solid-state targets used in 
scattering experiments [10, 13]. Coherent effects in spin systems, being 
similar to their coherent counterparts in optics [14, 15], could be used 
for analogous purposes but in another frequency region. For example, spin 
masers [16-18] can be realized. The sensitivity of the characteristic times 
of coherent relaxation to initial conditions could be used for creating 
ultrasensitive particle detectors [19].

\section{Electron-Nuclear Spin Hamiltonian}

To make our consideration applicable to a wide class of magnetic materials,
we take a rather general Hamiltonian 
\begin{equation}
\label{1}
\hat H=\hat H_e+\hat H_n+\hat H_{int}, 
\end{equation}
describing a realistic situation where the sample contains electrons with a
Hamiltonian $\hat H_e$ and nuclei with a Hamiltonian $\hat H_n$, their
interaction being given by $\hat H_{int}$. The electron Hamiltonian is 
\begin{equation}
\label{2}
\hat H_e=-\frac 12\sum\limits_{i\neq j}J_{ij}\vec S_i\cdot \vec
S_j-\mu _e\sum\limits_i\vec B\cdot \vec S_i, 
\end{equation}
where $J_{ij}$ is an exchange interaction; the indices $i,j=1,2,...,N_e$ 
here enumerate electrons; $\vec S_i$ is a spin operator; 
$\mu _e=g_e\mu _B$, with $g_e$ being the electronic gyromagnetic ratio and 
$\mu _B$, the Bohr magneton; $\vec B$ is a magnetic field. The nuclear Hamiltonian
has the form [20] commonly accepted in the theory of nuclear magnetic
resonance, 
\begin{equation}
\label{3}
\hat H_n=\frac 12\sum\limits_{i\neq j}
\sum\limits_{\alpha \beta}C_{ij}^{\alpha \beta }I_i^\alpha I_j^\beta -
\mu _n\sum\limits_i\vec B\cdot \vec I_i, 
\end{equation}
in which $i,j=1,2,...,N_n$ enumerate nuclei with dipole
interactions 
\begin{equation}
\label{4}
C_{ij}^{\alpha \beta }=\frac{\mu _n^2}{r_{ij}^3}(\delta _{\alpha\beta }
 - 3n_{ij}^\alpha n_{ij}^\beta ) 
\end{equation}
between each other, where $\mu _n=g_n\mu _N$, $g_n$ being the nuclear
gyromagnetic ratio, $\mu _N$, the nuclear magneton, and
$r_{ij}\equiv \left| \vec r_{ij}\right|, \;
\vec n_{ij}\equiv \vec r_{ij}/r_{ij},\; \vec r_{ij}\equiv\vec r_i-\vec r_j$;
the indices $\alpha $ and $\beta $ label the components of Cartesian vectors
($\alpha ,\beta =x,y,z$); $\vec I_i$ is a nuclear spin operator. The general
form of the hyperfine interactions [20, 21] between electrons and nuclei is 
\begin{equation}
\label{5}
\hat H_{int}=A\sum\limits_i\vec S_i\cdot \vec I_i+\frac
12\sum\limits_{i\neq j}\sum\limits_{\alpha \beta }A_{ij}^{\alpha \beta
}S_i^\alpha I_j^\beta , 
\end{equation}
containing an isotropic contact part with an interaction intensity $A$ and a
dipole part with the interactions 
\begin{equation}
\label{6}
A_{ij}^{\alpha \beta }=\frac{\mu _e\mu _n}{r_{ij}^3}(\delta_{\alpha \beta }
- 3n_{ij}^\alpha n_{ij}^\beta ). 
\end{equation}
The total magnetic field is the sum 
\begin{equation}
\label{7}
\vec B=H_0\vec e_z+H_1\vec e_x, \qquad H_1=H_a+H, 
\end{equation}
of an external magnetic field in the $z$ direction and of a transverse field
including an effective field of a transverse magnetocrystalline anisotropy
[22] and a feedback field $H$ of a resonator. The longitudinal part of the
magnetocrystalline anisotropy can be included into the external magnetic
field $H_0$.

In the preceding formulas we have used, for simplicity, the same indices, $i$
and $j$, to enumerate electrons and nuclei, keeping in mind that
for each particular case these indices run over different sets, so that for
electrons $i=1,2,...,N_e$ and for nuclei $i=1,2,...,N_n$. The corresponding 
electron and nuclear densities, $\rho_e\equiv N_e/V$ and $\rho_n\equiv
N_n/V$, where V is the volume of a sample, are, in general, different.

The resonator coil is directed along the $x$--axis, so that the current induced
in it is caused by the motion of the transverse magnetization 
\begin{equation}
\label{8}
M_x=\frac 1V\sum\limits_i(\mu _e<S_i^x>+\mu _n<I_i^x>), 
\end{equation}
where the angle brackets mean statistical averaging. The
resonant electric circuit is characterized by a natural frequency $\omega $,
ringing time $\gamma _3$, and quality factor $Q$, given by 
\begin{equation}
\label{9}
\omega \equiv \frac 1{\sqrt{LC}}, \qquad
\gamma _3\equiv \frac \omega{2Q}, \qquad
Q\equiv \frac{\omega L}R \; ,
\end{equation}
where $L$, $C$, and $R$ are inductance, capacity, and resistance,
respectively. The resonator feedback field is given [12] by the Kirchhoff
equation 
\begin{equation}
\label{10}
\frac{dH}{dt}+2\gamma _3H+\omega ^2\int\limits_0^tH(\tau )d\tau
=-4\pi \eta \frac{dM_x}{dt}, 
\end{equation}
in which $\eta $ is a filling factor.

\section{Coupled System of Equations}

Spin dynamics is defined by the Heisenberg equations for the electron spin
operators $S_i^{\pm }$ and $S_i^z$, and for the nuclear spin operators 
$I_i^{\pm}$ and $I_i^z$, where
$S_i^{\pm }=S_i^x\pm iS_i^y, \; I_i^{\pm }=I_i^x\pm iI_i^y$.
These equations are coupled to each other by hyperfine interactions
between electron and nuclear spins. Besides that, both types of spin motion
are coupled with the resonator through the resonator feedback field, defined
by equation (10), and the average transverse magnetization (8) expressed by
means of the average spins. To describe the dynamics of spins coupled with
each other as well as with a resonator, we need to derive the time evolution
equations for the average electron and nuclear spins 
\begin{equation}
\label{11}
x\equiv \frac 1{N_e}\sum\limits_i<S_i^{-}>, \qquad
z\equiv \frac 1{N_e}\sum\limits_i<S_i^z> \; ,
\end{equation}
\begin{equation}
\label{12}
u\equiv \frac 1{N_n}\sum\limits_i<I_i^{-}>, \qquad
s\equiv \frac 1{N_n}\sum\limits_i<I_i^z>. 
\end{equation}
The main steps of deriving these equations are the same as in Refs. [11, 12] 
except that now the Hamiltonian (1) is more complicated.

We introduce the Zeeman frequencies 
\begin{equation}
\label{13}
\omega _e\equiv \mu _eH_0, \qquad
\omega _n\equiv \mu _nH_0 
\end{equation}
and the anisotropy parameters 
\begin{equation}
\label{14}
\alpha _e\equiv \mu _eH_a, \qquad \alpha _n\equiv \mu _nH_a, 
\end{equation}
where the Planck constant is set $\hbar \equiv 1$. We use the notation
\begin{equation}
\label{15}
\xi _0\equiv \frac 12\sum\limits_{j(\neq i)}\left( \bar
a_{ij}<I_j^z>+\bar c_{ij}<I_j^{+}>+\bar c_{ij}^{*}<I_j^{-}>\right) , 
\end{equation}
\begin{equation}
\label{16}
\xi \equiv \frac 12\sum\limits_{j(\neq i)}\left( \bar
e_{ij}<I_j^{-}>+2\bar b_{ij}<I_j^{+}>+2\bar c_{ij}<I_j^z>\right) , 
\end{equation}
in which
$$
\overline a_{ij} \equiv A_{ij}^{zz} \; , \qquad 
\overline e_{ij} \equiv \frac{1}{2} \left ( A_{ij}^{xx} + A_{ij}^{yy}
\right ) \; ,
$$
$$
\overline b_{ij} \equiv \frac{1}{4} \left ( A_{ij}^{xx} - A_{ij}^{yy} -
2i A_{ij}^{xy}\right ) \; , \qquad
\overline c_{ij} \equiv \frac{1}{2} \left ( A_{ij}^{xz} - i A_{ij}^{yz}
\right ) \; ,
$$
$A_{ij}^{\alpha\beta}$ being given in Eq. (6). Also we write
$$
\varphi _0\equiv \sum\limits_{j(\neq i)}\left[
(a_{ij}-e_{ij})<I_j^z>+c_{ij}<I_j^{+}>+c_{ij}^{*}<I_j^{-}>+ \right.
$$
\begin{equation}
\label{17}
+ \left. \frac 12\left( \bar a_{ij}<S_j^z>+\bar c_{ij}<S_j^{+}>+\bar
c_{ij}^{*}<S_j^{-}>\right) \right] , 
\end{equation}
$$
\varphi \equiv \sum\limits_{j(\neq i)}\left[ 2\left(
b_{ij}<I_j^{+}>+c_{ij}<I_j^z>\right) + \right.
$$
\begin{equation}.
\label{18}
+ \left. \frac 12\left( \bar
e_{ij}<S_j^{-}>+2\bar b_{ij}<S_j^{+}>+2\bar c_{ij}<S_j^z>\right) 
\right]\; ,
\end{equation}
where
$$
a_{ij} \equiv C_{ij}^{zz} \; , \qquad 
e_{ij} \equiv \frac{1}{2} \left ( C_{ij}^{xx} + C_{ij}^{yy}\right ) \; ,
$$
$$
b_{ij} \equiv \frac{1}{4} \left ( C_{ij}^{xx} - C_{ij}^{yy} -
2i C_{ij}^{xy} \right ) \; , \qquad
c_{ij} \equiv \frac{1}{2} \left ( C_{ij}^{xz} - i C_{ij}^{yz}\right )\; ,
$$
$C_{ij}^{\alpha\beta}$ being defined in Eq. (4).
Equations (15) to (18) describe local random fields caused by spin fluctuations.
In the uniform approximation, all these quantities would be zero, because of
the properties of dipole interactions. However, these local fields cannot be
neglected, since they play a crucial role at the initial stage of spin
relaxation. Therefore they must be retained and treated as local random
variables. For other terms in the evolution equations, having a long-range nature in real space, one may employ the semiclassical approximation. This
approach of using for long-range terms the semiclassical approximation
complemented by the stochastic quantization of short-range terms has been
developed in Refs [11, 12].

Following these steps and taking into account the longitudinal, 
$\gamma_1,\Gamma_1$, and transverse, $\gamma_2,\;\Gamma_2$, attenuations 
for the electron and nuclei, respectively, we obtain the evolution
equations for the electron spin variables (11) and nuclear spin variables
(12). For the transverse and longitudinal electron spins we have, 
respectively,
\begin{equation}
\label{19}
\frac{dx}{dt}=i\left( \omega _e+i\gamma _2-\xi _0-As\right)
x-i\left( \alpha _e+\mu _eH-\xi -Au\right) z, 
\end{equation}
\begin{equation}
\label{20}
\frac{dz}{dt}=\frac i2\left( \alpha _e+\mu _eH-\xi -Au\right)
x^{*}-\frac i2\left( \alpha _e+\mu _eH-\xi ^{*}-Au^{*}\right) x-\gamma
_1\left( z-\sigma \right) , 
\end{equation}
where $\gamma_1$ and $\gamma_2$ are attenuation parameters and $\sigma$
is the stationary value of $z$. The variable $x$ is complex,
while $z$ is real, so we should add either an
equation for $x^{*}$ or for $\left| x\right| $. It is convenient to
consider 
\begin{equation}
\label{21}
\frac{d\left| x\right| ^2}{dt}=-2\gamma _2\left| x\right|
^2+i\left( \alpha _e+\mu _eH-\xi ^{*}-Au^{*}\right) zx-i\left( \alpha _e+\mu
_eH-\xi -Au\right) zx^{*}. 
\end{equation}
In the case of nuclear spins, we obtain 
\begin{equation}
\label{22}
\frac{du}{dt}=i\left( \omega _n+i\Gamma _2-\varphi _0-Az\right)
u-i\left( \alpha _n+\mu _nH-\varphi -Ax\right) s 
\end{equation}
\begin{equation}
\label{23}
\frac{ds}{dt}=\frac i2\left( \alpha _n+\mu _nH-\varphi -Ax\right)
u^{*}-\frac i2\left( \alpha _n+\mu _nH-\varphi ^{*}-Ax^{*}\right) u-\Gamma
_1\left( s-\varsigma \right) , 
\end{equation}
where $\Gamma _1$ and $\Gamma _2$ are the longitudinal and transverse
attenuations, respectively, and $\varsigma $ is the stationary value of $s$. In
addition, we shall need

\begin{equation}
\label{24}
\frac{d\left| u\right| ^2}{dt}=-2\Gamma _2\left| u\right|
^2+i\left( \alpha _n+\mu _nH-\varphi ^{*}-Ax^{*}\right) su-i\left( \alpha
_n+\mu _nH-\varphi -Ax\right) su^{*}. 
\end{equation}

All equations (19) to (24) contain the resonator feedback field H described
by (10). The latter can be transformed [12] to the
integral feedback equation 
\begin{equation}
\label{25}
H=-4\pi \eta \int\limits_0^tG(t-\tau )dM_x(\tau ), 
\end{equation}
expressed through a Stieltjes integral with the Green function
$$
G(t)=(\cos \omega _3t-\frac{\gamma _3}{\omega _3}\sin \omega _3t)
\exp(-\gamma _3t )
$$
and the differential measure $dM_x$ with $M_x$ defined in Eq. (8). 
Here the effective frequency is
$\omega _3\equiv \sqrt{\omega ^2-\gamma _3^2}$.

The system of seven nonlinear equations (19) to (25) determines the dynamics
of electron and nuclear spins coupled with each other as well as with a
resonator.

\section{Scale Separation Approach}

Our aim here is to study the strongly nonequilibrium regimes of spin motion.
This problem is different from considering the equilibrium
properties of coupled electron and nuclear spins [23-26]. An
additional complication, in our case, arises from the coupling of spins with
a resonator by means of the feedback equation (25). To solve equations (19) to (25), we employ the scale separation approach [11, 12]
which is a generalization of the averaging techniques of dynamical theory [27,
28] to statistical systems.

To understand what different time scales exist for the system considered, we
need to specify what small parameters we have. Since we have a
sample coupled with a resonator, some small parameters should appear by
concretizing the corresponding resonance conditions, assuming the ringing width is
much smaller than the natural frequency, 
\begin{equation}
\label{26}
\frac{\gamma _3}\omega \ll 1 .
\end{equation}
The resonator natural frequency can be tuned either to the frequency of
electron spin resonance $\omega _e$, so that 
\begin{equation}
\label{27}
\left| \frac{\Delta _e}{\omega _e}\right| \ll 1, \qquad
\Delta _e\equiv \omega -\omega _e, 
\end{equation}
or to the frequency of nuclear magnetic resonance 
\begin{equation}
\label{28}
\omega _N\equiv \omega _n - A z_{eff}, 
\end{equation}
in which $z_{eff}$ is the longitudinal electron spin $z$ averaged over the
period $2\pi /\omega $, so that
\begin{equation}
\label{29}
\left| \frac{\Delta _N}{\omega _N}\right| \ll 1, \qquad
\Delta _N\equiv \omega -\omega _N. 
\end{equation}

We assume that the external magnetic field $H_0$ is sufficiently
strong that 
\begin{equation}
\label{30}
\left| \frac{\alpha _e}{\omega _e}\right| \ll 1, \qquad
\left| \frac{\mu_eH_{eff}}{\omega _e}\right| \ll 1, \qquad
\left| \frac A{\omega _e}\right| \ll 1, 
\end{equation}
where $H_{eff}$ is the resonator feedback field averaged over a period
$2\pi /\omega $. We also assume 
\begin{equation}
\label{31}
\left| \frac{\gamma _1}{\omega _e}\right| \ll 1, \qquad
\left| \frac{\gamma_2}{\omega _e}\right| \ll 1. 
\end{equation}
Then from Eqs. (19) to (21) it follows that the variables $z$ and $\left|
x\right| ^2$ are to be treated as slow compared to the fast variable $x$.
Similarly, for nuclei we assume 
\begin{equation}
\label{32}
\left| \frac{\alpha _n}{\omega _N}\right| \ll 1, \qquad
\left| \frac{\mu_nH_{eff}}{\omega _N}\right| \ll 1, \qquad
\left| \frac{Ax_{eff}}{\omega _N}\right| \ll 1, 
\end{equation}
where the subscript {\it eff} means again that the corresponding quantity
is
averaged over $2\pi /\omega $, and we keep in mind the usual inequalities 
\begin{equation}
\label{33}
\left| \frac{\Gamma _1}{\omega _N}\right| \ll 1, \qquad
\left| \frac{\Gamma_2}{\omega _N}\right| \ll 1. 
\end{equation}
Then Eqs. (22), (23), and (24) show that the variables $s$ and $\left|
u\right| ^2$ are slow compared to the fast variable $u$. As the
nuclear magnetic moment $\mu _n$ is much smaller than that of an electron, $\mu
_e$, we have the inequalities 
\begin{equation}
\label{34}
\left| \frac{\mu _n}{\mu _e}\right| \ll 1, \qquad
\frac{\Gamma _1}{\gamma _1} \ll 1, \qquad
\frac{\Gamma _2}{\gamma _2}\ll 1. 
\end{equation}
Then, comparing Eqs. (19) and (22), we see that the variable $u$ is slow compared to the fast variable $x$. And the comparison of Eqs. (21) and (24)
tells us that $\left| u\right| ^2$ is slow compared to the faster $\left|
x\right| ^2$.

One more condition assumed is related to the local random fields (15) to
(18). These local fields define the parameters of inhomogeneous broadening
due to the electron-nuclear, $\Gamma_{en}$, and nuclear-nuclear, $\Gamma _{nn}$, 
interactions. These widths are assumed to satisfy
\begin{equation}
\label{35}
\left| \frac{\Gamma _{en}}{\omega _N}\right| \ll 1, \qquad
\left| \frac{\Gamma _{nn}}{\omega _N}\right| \ll 1. 
\end{equation}

Using conditions (30) and (32), we may simplify the feedback equation (25)
to 
\begin{equation}
\label{36}
H= - 2{\rm Re} \left( \beta _e\frac{dx}{dt}+\beta _n\frac{du}{dt}\right), 
\end{equation}
in which the parameters 
\begin{equation}
\label{37}
\beta _e\equiv \pi ^2\eta \frac{\mu _e\rho_e}\omega , \qquad
\beta _n\equiv \pi ^2\eta \frac{\mu _n\rho_n}\omega 
\end{equation}
characterize the effective coupling of the sample with the resonator.
The details are given in the Appendix. Substituting into Eq.(36) the
derivatives from (19) and (22), we find 
$$
H=- 2{\rm Re}\; i\beta _e\left[ \left( \omega _e-As-\xi _0+i\gamma
_2\right) x+\left( Au+\xi \right) z\right] -
$$
\begin{equation}
\label{38}
-2{\rm Re}\; i\beta _n\left[ \left(
\omega _n-Az-\varphi _0+i\Gamma _2\right) u+\left( Ax+\varphi \right)
s\right] . 
\end{equation}
The feedback field (38) is to be substituted into equations (19) to (24).

Then we solve Eq. (19) treating there all slow variables as quasi-integrals
of motion. The solution reads
\begin{equation}
\label{39}x=\left( x_0-\bar x\right) \exp \left\{ \left( i\bar \Omega
_e-\bar \gamma _2\right) t\right\} +\bar x, 
\end{equation}
where $x_0=x(0)$ and
$$
\bar \Omega _e\equiv \omega _e-As-\xi _0-\mu _e\beta _e\gamma _2z, \qquad
\bar \gamma _2\equiv \gamma _2+\mu _e\beta _e\left( \omega_e
- As - \xi_0\right ) z, 
$$
$$
\bar x\equiv \frac 1{\bar \Omega _e}\left\{ \left( \alpha _e-Au-\xi \right)
z+i\mu _e\beta _e\left[ A\left( u^{*}-u\right) +\xi ^{*}-\xi \right]
z^2\right\} . 
$$
After this, we solve Eq. (22), keeping the slow variables fixed and
averaging the fast variables $x(t)$ and $z(t)$ to obtain $x_{eff}$ and $z_{eff}$.
Since $x(t)$ is already known, we have
\begin{eqnarray}
x_{eff}=\left\{ \begin{array}{cc}
A_ez/\Omega _e, & \omega \approx \omega_e \\
\nonumber
0 , & \omega \approx \omega _N\; ,
\end{array} \right.
\end{eqnarray}
where $\Omega_e\equiv\omega_e - As -\xi_0$ and $A_e\equiv\alpha_e - Au-\xi$.
The solution of Eq. (22) is 
\begin{equation}
\label{40}
u=\left( u_0-\bar u\right) \exp \left\{ \left( i\bar \Omega_n-
\bar \Gamma _2\right) t\right\} +\bar u, 
\end{equation}
where $u_0\equiv u(0)$ and
$$
\bar \Omega _n\equiv \omega _n-A\left( 1+\mu _n\beta _es\right)
z_{eff}-\varphi _0-\mu _n\beta _n\Gamma _2s, 
$$
$$
\bar \Gamma _2\equiv \Gamma _2+\mu _n\beta _n\left( \omega_n -
Az_{eff}-\varphi _0\right) s, 
$$
$$
\bar u\equiv \frac 1{\bar \Omega _n}\left[ \left( \alpha _n-\varphi \right)
s+i\mu _n\beta _e\left( \xi ^{*}-\xi \right) sz_{eff}\right] . 
$$

The solutions (39) and (40) are to be substituted into the equations
for the slow variables, with the right-hand sides of the latter equations being
averaged over time and over random local fields according to the rule
$$
\ll f(t,\psi)\gg \; = \; \int \left [ \frac{\omega}{2\pi} 
\int_0^{2\pi/\omega} f(t,\psi)\; dt \right ] \; d \; m(\psi) \; ,
$$ 
with the stochastic measure $m(\psi)$ such that
$$
\ll \xi_0 \gg \; = \; \ll \xi\gg \; = \; \ll \xi_0\xi \gg \; = 0 \; ,
$$
$$
\ll \varphi_0 \gg \; = \; \ll \varphi\gg \; = \; 
\ll \varphi_0\varphi \gg \; = 0 \; ,
$$
$$
\ll \xi_0\varphi_0 \gg \; = \; \ll \xi_0\varphi \gg \; = \; 
\ll \xi\varphi_0  \gg \; =\; \ll \xi\varphi \gg \; = 0 \; , 
$$
$$
\ll \xi_0^2 \gg \; = \; \ll |\xi|^2\gg \; = \gamma_*^2 = \Gamma_{en}^2 \; ,
$$
$$
\ll \varphi_0^2 \gg \; = \; \ll |\varphi|^2\gg \; = \; \Gamma_*^2=
\Gamma_{nn}^2 + \Gamma_{en}^2 \; .
$$
The constants $\gamma_*$ and $\Gamma_*$ are the parameters of 
inhomogeneous broadening caused by hyperfine electron-nuclear dipole 
interactions and by nuclear dipole interactions.

We introduce the effective coupling parameters 
\begin{equation}
\label{41}
g_e\equiv \pi ^2\eta \frac{\rho_e\mu _e^2\omega _E}{\gamma _2\omega }
\left( 1+\frac{\rho_n\mu _nAs}{\rho_e\mu _e\omega _E}\right) , 
\end{equation}
\begin{equation}
\label{42}
g_n\equiv \pi ^2\eta \frac{\rho_n\mu _n^2\omega _N}{\Gamma _2\omega }
\left( 1+\frac{\rho_e\mu _eAz_{eff}}{\rho_n\mu _n\omega _N}\right) , 
\end{equation}
characterizing the strength of coupling between electron or nuclear spins,
respectively, and the resonator. The effective frequencies are 
\begin{equation}
\label{43}
\omega _E\equiv \omega _e-As, \qquad
\omega _N\equiv \omega _n-Az_{eff}, 
\end{equation}
which are the frequencies of the electron spin resonance and nuclear
magnetic resonance.

Also, we define new slow variables for electrons, 
\begin{equation}
\label{44}
v=\left| x\right| ^2-\frac{\alpha _e^2+A^2\left| u\right|
^2+\gamma _{*}^2}{\omega _E^2}z^2, 
\end{equation}
and for nuclei 
\begin{equation}
\label{45}w=\left| u\right| ^2-\frac{\alpha _n^2+\Gamma _{*}^2+\delta ^2}{%
\omega _N^2}s^2, 
\end{equation}
where

\begin{equation}
\label{46}
\delta \equiv \sqrt{2}\pi ^2\eta \gamma _{*}
\frac{\rho_e\mu _e\mu _n}{\omega _N}z_{eff}. 
\end{equation}

Accomplishing all these steps, we obtain from Eqs. (20) to (24)
\begin{equation}
\label{47}\frac{dz}{dt}=\gamma _2g_ev-\gamma _1(z-\sigma ), 
\end{equation}
\begin{equation}
\label{48}\frac{dv}{dt}=-2\gamma _2(1+g_ez)v, 
\end{equation}
\begin{equation}
\label{49}\frac{ds}{dt}=\Gamma _2g_nw-\Gamma _1(s-\zeta ), 
\end{equation}
\begin{equation}
\label{50}\frac{dw}{dt}=-2\Gamma _2(1+g_ns)w. 
\end{equation}
Equations (47) to (50) define the averaged motion of slow variables. We require their solutions, since all observable quantities
can be expressed through them.

\section{Nuclear Spin Dynamics}

Equations (47) to (50) show that the electron spin dynamics is qualitatively
similar to that of nuclear spins, but there are three main points
distinguishing electron from nuclear relaxation. First, electron spin
processes are usually much faster than nuclear processes [3], which is
related to the fact that $\gamma _2\gg \Gamma _2$. Second, as the electronic
magnetic moment is three orders of magnitude larger than the nuclear magneton,
electron spin motion is much less influenced by the presence of nuclei than
the motion of nuclear spins by the existence of electrons. Third, the
stronger influence of electrons on the motion of nuclear spins is caused
by a long--range magnetic order that more readily occurs in
electronic systems than in nuclear ones. Therefore nuclear spin dynamics
is a little more complicated but at the same time richer than the dynamics
of electron spins.

Suppose that the electron spins either were not perturbed at the initial time 
or, if perturbed, that fast electron processes have already been relaxed to
their stationary state. Let us study the dynamics of nuclear spins that were
initially prepared in a strongly nonequilibrium state. We denote the initial
conditions for the nuclear spin variables as
\begin{equation}
\label{51}
s(0)=s_0, \qquad w(0)=w_0, \qquad u(0)=u_0. 
\end{equation}

If relaxation of nuclear spins lasts for times of order $T_1\equiv
\Gamma _1^{-1}$, this would mean that coherent processes do not develop.
Such a case would be of no interest for us, since our aim here is to
investigate the fast coherent relaxation. Hence we shall consider times
such that $t\ll\Gamma _1^{-1}$. In this case, we may omit in Eq. (49) the term
containing $\Gamma _1$. Then we can solve Eqs. (49) and (50) analytically
obtaining for the longitudinal nuclear spin 
\begin{equation}
\label{52}
s=\frac{\gamma _0}{g\Gamma _2}\tanh \left( \frac{t-t_0}{\tau _0}
\right) -\frac 1g 
\end{equation}
where, for the sake of simplicity, we write $g\equiv g_n$, and 
\begin{equation}
\label{53}
w=\left( \frac{\gamma _0}{g\Gamma _2}\right) ^2{\rm sech}^2\left( 
\frac{t-t_0}{\tau _0}\right) . 
\end{equation}
Here $\gamma _0\equiv \tau _0^{-1}$ is the relaxation width given by the
equation 
\begin{equation}
\label{54}
\gamma _0=\Gamma _2\left[ \left( 1+gs_0\right) ^2+g^2w_0\right]^{1/2} ,
\end{equation}
with $\tau _0$ being the relaxation time, and $t_0$ the delay time, 
\begin{equation}
\label{55}t_0=\frac{\tau _0}2\ln \left| \frac{\gamma _0-\Gamma _2\left(
1+gs_0\right) }{\gamma _0+\Gamma _2\left( 1+gs_0\right) }\right| . 
\end{equation}
According to Eq. (45), the modulus squared of the transverse nuclear spin is
\begin{equation}
\label{56}
\left| u\right| ^2=\left( \frac{\gamma _0}{g\Gamma _2}\right)^2
{\rm sech}^2\left( \frac{t-t_0}{\tau _0}\right) +\frac{\alpha _n^2+\Gamma
_{*}^2+\delta ^2}{\omega _N^2}s^2. 
\end{equation}

If the coupling parameter $g$, is small, $g\ll 1 $, then Eqs. (49) and (50) 
show that the relaxation of nuclear spins follows the standard exponential 
law with the relaxation times $T_1\equiv\Gamma_1^{-1}$ and  
$T_2\equiv\Gamma _2^{-1}$. This trivial regime is not interesting for us, 
so we concentrate attention on the case of strong coupling, when $g>>1$. 
Then the relaxation width (54) can be written as 
\begin{equation}
\label{57}
\gamma _0=\Gamma _2\left( g\sqrt{s_0^2+w_0}+
\frac{s_0}{\sqrt{s_0^2+w_0}}\right) . 
\end{equation}
For the delay time (55), we find 
\begin{equation}
\label{58}
t_0=\frac{\tau _0}2\ln \left| \frac{g\left( s_0^2+w_0\right)
+s_0-\left( 1+gs_0\right) \sqrt{s_0^2+w_0}}{g\left( s_0^2+w_0\right)
+s_0+\left( 1+gs_0\right) \sqrt{s_0^2+w_0}}\right| . 
\end{equation}
For the relaxation time, after using again the inequality $g\gg 1$, we
have
\begin{equation}
\label{59}
\tau _0=\frac{T_2}{g\sqrt{s_0^2+w_0}}. 
\end{equation}
A large value of the coupling constant $g$ means, according to its 
definition in Eq. (42), that nuclear spins are strongly correlated with 
each other by means of an effective interaction through the resonator 
feedback field. As a result of this correlation they move coherently, 
which leads to the nonzero value of $w\approx |u|^2$. Recall that, by 
definition (12), $u=0$ for incoherent spins. Coherent motion of the spins 
results in their ultrafast relaxation, which follows from Eq. (59) 
yielding $\tau_0\ll T_2$ when $g\gg 1$. This is why the characteristic 
time (59) can be called the {\it coherent relaxation time}. Notice also 
that the coupling parameter (42) is proportional to the number of nuclei 
$N_n$, and so $\tau_0\sim 1/g\sim 1/N_n$. Such a dependence of the 
relaxation time, $\tau_0\sim 1/N_n$, on the number of radiators is 
typical for coherent processes that in optics are called superradiance [14, 
15].

The relaxation characteristics, as is seen, essentially depend on the
initial conditions (51). If $u_0\neq 0$ at the initial time, this implies
that an initial coherence is imposed on the spins, which can be done by means of a
short external pulse of a transverse field. When $u_0\sim 1$, then we have 
$w_0\approx\left|u_0\right|^2$, since the second term in Eq. (45) is small. 
In this case we get the regime of triggered relaxation [12].

A much more interesting question concerns how the relaxation of nuclear
spins starts when no initial coherence is thrust upon the spins. This
problem is also more important than consideration of the case when
relaxation is triggered by external fields. For, when the relaxation is
initiated not by external forces but by internal interactions, the
relaxation characteristics, such as the delay time and the relaxation
time, significantly depend on the parameters of these internal
interactions. Therefore, the self--organized relaxation
reflects (and provides 
information about) the intrinsic properties of matter.

Let us consider the {\it self-organized relaxation} in the system of nuclear 
spins strongly coupled with a resonator. That is, we analyze the case
corresponding to the conditions: 
\begin{equation}
\label{60}
u_0=0, \qquad s_0\neq 0, \qquad g\left| s_0\right| >>1. 
\end{equation}
If $u_0=0$, then, according to relation (45), we have $\left|w_0\right|\ll
s_0^2$. This permits us to simplify the relaxation width (57) getting 
\begin{equation}
\label{61}\gamma _0=\frac{\Gamma _2}{2s_0^2}\left[ g\left| s_0\right| \left(
2s_0^2+w_0\right) +\varepsilon \left( 2s_0^2-w_0\right) \right] , 
\end{equation}
where $\varepsilon \equiv $ sgn $s_0$. The delay time (58) becomes 
\begin{equation}
\label{62}
t_0=\frac{\tau _0}2\ln \left| \frac{2\left( 1-\varepsilon \right)
s_0^2+w_0}{2\left( 1+\varepsilon \right) s_0^2+w_0}\right| ,
\end{equation}
and for the relaxation time (59), we get

\begin{equation}
\label{63}\tau _0=\frac{2s_0^2-w_0}{2g\left| s_0\right| ^3\Gamma _2}\simeq 
\frac{T_2}{g\left| s_0\right| }. 
\end{equation}

The delay time (62) strongly depends on the sign of the initial polarization
of nuclear spins, $\varepsilon \equiv $ sgn $s_0$. When this initial
polarization is directed along the external magnetic field $H_0$, i.e.,
along the $z$--axis, then 
\begin{equation}
\label{64}
t_0=\frac{\tau _0}2\ln \left| \frac{w_0}{4s_0^2}\right| \qquad
\left(\varepsilon =1\right) , 
\end{equation}
and if the initial polarization is directed opposite to the external field, then
\begin{equation}
\label{65}
t_0=\frac{\tau _0}2\ln \left| \frac{4s_0^2}{w_0}\right| \qquad
\left( \varepsilon =-1\right) . 
\end{equation}

In the case when the initial polarization is along the $z$--axis, Eq. (64)
shows that $t_0<0$, since $\left| w_0\right|\ll s_0^2$. Then the function
(53) quickly decreases starting from $t=0$. This function $w(t)$, being
proportional to $\left| u\right| ^2$, describes the degree of coherence
in the motion of nuclear spins. In turn, the function $\left|
u\right| ^2$ is proportional to the power flow in the resonator
circuit and, thus, is a directly measurable quantity [11, 12]. Hence, when
$w(t)$ quickly decreases starting from the initial time $t=0$ and
$\left| w_0\right|\ll s_0^2$, this
means that no noticeable coherence develops in the system.

In contrast, if the initial polarization of nuclear spins is directed
opposite to the external magnetic field, so that the system is prepared in a
strongly nonequilibrium state, then, from Eq. (65), there is a
positive solution for the delay time $t_0>0$. In such a case, the function
(53) increases from its initial value $w_0$, reaching a maximum at 
$t=t_0$, when 
\begin{equation}
\label{66}
w\left( t_0\right) =s_0^2, \qquad s\left( t_0\right) =-\frac 1g. 
\end{equation}
This means that a self-organized coherent pulse develops with a maximum at $%
t=t_0$, which explains why $t_0$ is called the delay time.

Combining both the cases, (64) and (65), into one and substituting the expressions 
for $\tau _0$ and $w_0$, we obtain 
\begin{equation}
\label{67}t_0=\frac{T_2}{2gs_0}\ln \left| \frac{\alpha _n^2+\Gamma
_{en}^2+\Gamma _{nn}^2+\delta ^2}{4\omega _N^2}\right| . 
\end{equation}
This is the central formula for analyzing what internal microscopic
mechanisms are responsible for the self-organized development of coherent
relaxation. Each of these internal mechanisms is related to the
corresponding parameter entering formula (67). Among such internal causes
that may trigger self-organized coherence, we have the transverse
magnetocrystalline anisotropy, $\alpha _n$; the dipole part of the hyperfine
interactions, characterized by the parameter $\Gamma _{en}$; the dipole
interactions between nuclear spins, $\Gamma _{nn}$; and the parameter $%
\delta $ defined in Eq. (46), which is due to the simultaneous existence of
the hyperfine interactions, of coupling with a resonator, and of the
magnetization of electron spins.

The relaxation time (54), for the case of self-organized relaxation,
when $u_0=0$, reads 
\begin{equation}
\label{68}
\tau _0=T_2\left[ \left( 1+gs_0\right) ^2 + g^2s_0^2\; 
\frac{\alpha_n^2+\Gamma _{en}^2+\Gamma _{nn}^2+\delta ^2}{\omega _N^2}
\right] ^{-1/2}. 
\end{equation}
This demonstrates that the value of the relaxation time (68) depends mainly
on the strength, $g$, of coupling with a resonator. In this way, the delay 
time (67) and the relaxation time (68) are related to different
characteristics of the system considered.

In order to decide what kind of interactions, direct dipole interactions
between nuclei or hyperfine interactions, influences more the values of the
characteristic times, we should compare the corresponding widths 
$\Gamma_{nn}\sim \rho_n\mu _n^2$ and $\Gamma _{en}\sim \rho\mu _e\mu _n$,
where $\rho\approx\min \left\{\rho_e,\rho_n\right\} $. There are two
limiting cases. The first is when $\rho_n\leq \rho_e$, and then 
\begin{equation}
\label{69}
\frac{\Gamma _{nn}}{\Gamma _{en}}\sim \frac{\mu _n}{\mu _e}\sim 10^{-3} 
\qquad \left( \rho_n\leq \rho_e\right) . 
\end{equation}
Hence, when the density of nuclei is lower or comparable with that of
electrons, nuclear dipole interactions are negligibly small compared to
the hyperfine interactions between nuclei and electrons. Another case is
when the density of nuclei is much higher than that of electrons; then 
\begin{equation}
\label{70}
\frac{\Gamma _{nn}}{\Gamma _{en}}\sim \frac{\rho_n\mu _n}{\rho_e\mu _e}
\sim \frac{\rho_n}{\rho_e}10^{-3} \qquad \left( \rho_n \gg \rho_e\right) . 
\end{equation}
Thus, the nuclear dipole interactions become stronger than the
hyperfine interactions only when the density of nuclei surpasses by three
orders of magnitude the density of electrons.

In equilibrium theory, the influence of hyperfine interactions is often
modelled by the effective Suhl-Nakamura forces directly acting between
nuclear spins [21]. These forces are responsible for the appearance of
nuclear spin waves corresponding to well defined excitations, even at those
temperatures where the nuclear spins are completely disordered. The
underlying cause of the formation of the nuclear spin waves is the existence
of magnetic long-range order in the electronic subsystem, which defines both
the long-range interaction radius of the Suhl-Nakamura force and its
existence as such. The effective Suhl-Nakamura force describes an indirect
interaction of nuclear spins through magnetically ordered electrons [21].

For strongly nonequilibrium processes, such as those considered in this paper, the role
of the magnetic order of the electrons is essentially different. This order does
strongly influence several important characteristics. For instance, in
addition to the usual shift of the nuclear magnetic resonance frequency $%
\omega _N=\omega _n-Az_{eff}$, it leads to the appearance of parameter
(46) playing for nuclei the role of an additional inhomogeneous width.
Nevertheless, even if this order is absent, so that $z_{eff}\rightarrow 0$,
coherent nuclear spin relaxation can exist. When $z_{eff}\rightarrow 0$, the
Suhl-Nakamura force is not well defined, but the hyperfine interactions do
not stop existing. These interactions define the width $\Gamma _{en}$, which
is not zero even if the electronic magnetic order is absent. Thus, the
presence of hyperfine interactions is already important, even when there is
no long-range magnetic order, when the Suhl-Nakamura force and nuclear spin
waves are not well defined.

However, it is worth emphasizing that the appearance of electronic magnetic
order can strongly change the values of the characteristic parameters. Thus,
one of the most important parameters is the effective coupling (42)
describing the coupling of nuclear spins with a resonator. The value of this
parameter essentially depends on whether $z_{eff}$ is zero or not. The
appearance of magnetic order in the electronic system can change the value
of the parameter (42) by several orders of magnitude which, in turn, drastically changes the
values of the delay time $t_0$ and the relaxation time $\tau _0$.

\section{Characteristic Numerical Values}

To understand better the role of different factors in the coherent
relaxation of nuclear spins and the magnitudes of the related characteristic
parameters, let us now make numerical estimates. We take the values of
parameters that are typical of many ferromagnetic materials in
which one usually studies nuclear magnetic resonance and nuclear spin echo
[21, 29, 30]. These can be pure materials, such as Co, or various
ferromagnetic alloys and compounds [29, 30]. Since ferrimagnets are often
treated by effective ferromagnetic models, ferrimagnetic materials, such as
MnFe$_2$O$_3$, are also included here [31].

For the characteristic magnetic fields and the corresponding frequencies we
have the following values. The contact hyperfine field $H_A\equiv A/\mu
_n\sim 10^5$ G, the related frequency $\omega _A\equiv A/\hbar \sim 
10^9$ s$^{-1}$. The hyperfine field is smaller than the electron exchange 
field $H_J\equiv J/\mu _e\sim 10^6$ G, the corresponding frequency being 
$\omega_J\equiv \mu _eH_J/\hbar \sim 10^{13}$ s$^{-1}$. However, both these
fields are important for nonequilibrium processes in the nuclear spin
system, although the role of these fields is different. The hyperfine field
acts directly on the nuclear spins, and the exchange field influences nuclear
spin relaxation indirectly, through the formation of magnetic order in the
electronic subsystem. If we take an external magnetic field $H_0\sim 10^4$ G,
then the Zeeman frequencies (13) are $\omega _e\equiv \mu _eH_0/\hbar \sim
10^{11}$ s$^{-1}$ and $\omega _n\equiv \mu _nH_0/\hbar \sim 10^8$ s$^{-1}$. 
The magnetocrystalline anisotropy field $H_a\leq 10^3$ G,
depending on the particular structure of matter. The
anisotropy parameters (14) are $\alpha _e\equiv \mu _eH_a/\hbar \leq 10^{10}$
s$^{-1}$ and $\alpha _n\equiv \mu _nH_a/\hbar \leq 10^7$ s$^{-1}$, respectively. The 
longitudinal widths $\gamma _1$ and $\Gamma _1$ can vary within rather
wide intervals, but usually $\gamma _1\ll\gamma _2$ and $\Gamma_1\ll\Gamma_2$.
For the transverse widths we may take the estimates $\gamma _2\sim \rho_e
\mu_e^2/\hbar $ and $\Gamma_2\sim \rho_n\mu _n^2/\hbar$. This, with 
$\mu _e\sim 10^{-20}$ erg/G, $\mu_n\sim 10^{-23}$ erg/G, and $\rho_e\sim
\rho_n\sim 10^{23}$ cm$^{-3}$, gives $\gamma_2\sim 10^{10}$ s$^{-1}$ and 
$\Gamma_2\sim 10^4$ s$^{-1}$. The value for $\gamma_2$ is to be treated as
the upper limit, since the density of electrons is usually less than 
10$^{23}$ cm$^{-3}$, being, for instance, 10$^{22}$ cm$^{-3}$ for typical 
ordinary metals [32]. In the case when the considered
electrons are related to impurity ions inside an insulator, as in Refs
[6-10], then their density can be $\rho_e\sim 10^{20}$ cm$^{-3}$,
resulting in $\gamma _2\sim 10^7$s$^{-1}$. The estimated value of 
$\Gamma _2$ is in agreement with experimental measurements [30]. For the
resonator ringing width, we may take a typical experimental value
of $\gamma _3\sim 10^6$ s$^{-1}$. Then $\gamma _3/\omega
_e\sim 10^{-5}$ and $\gamma _3/\omega _n\sim 10^{-2}$, so that inequality
(26) is satisfied.

Since $\gamma _1\ll\gamma _2$ and $\gamma _2/\omega _e\leq 10^{-1}$,
condition (31) is valid. The nuclear magnetic resonance frequency (28) is 
$\omega _N\sim \omega _n\sim 10^8$ s$^{-1}$ if $z_{eff}=0$, that is if the
magnetic order is absent, and if $z_{eff}\neq 0$, then 
$\omega _N\sim 10^9$  s$^{-1}$. Hence, $\Gamma_2/\omega_N\sim\Gamma_2/
\omega_n\sim 10^{-4}$, when $z_{eff}=0$, and $\Gamma _2/\omega _N\sim
10^{-5}$ for a ferromagnetic material with $z_{eff}\neq 0$. This, together
with $\Gamma_1\ll\Gamma _2$, shows that condition (33) holds true.

For electrons, $\alpha _e/\omega _e\leq 10^{-1}$ and $\omega _A/\omega
_e\sim 10^{-2}$, while for nuclei, $\alpha _n/\omega _N\leq 10^{-2}$ and 
$x_{eff}=0$. The resonator feedback field (38) is $H_{eff}\sim\beta_e\omega
_A$. For electrons, with a resonator natural frequency close to the electron
spin resonance frequency, $\omega \sim \omega _e$, we have $\mu
_eH_{eff}/\omega _e\sim 10^{-3}$, and for nuclei, when $\omega \sim \omega
_N $, we find $\mu _nH_{eff}/\omega _N\sim 10^{-2}$. Thus, all inequalities
in equations (30) and (32) are valid. Since $\left| \mu _n/\mu _e\right|
\sim 10^{-3}$ and $\Gamma _1/\gamma _1$ and $\Gamma _2/\gamma _2$ are of
the order of $\rho_n\mu _n^2/\rho_e\mu _e^2\sim \left( \rho_n/\rho_e\right)
10^{-6}$, the inequalities in Eq. (34) hold true if $\rho_n$ and $\rho_e$ are
not drastically different. Conditions (35) are also satisfied, since 
$\Gamma_{en}/\omega _N\sim \mu_n\gamma _2/\mu _e\omega _N\sim 10^{-2}$ and 
$\Gamma _{nn}/\omega _N\sim \Gamma _2/\omega _N\sim 10^{-5}$.

Among the parameters defining the characteristic times of the coherent
nuclear spin relaxation, we have the anisotropy parameter $\alpha _n\leq
10^7 $ s$^{-1}$, the inhomogeneous broadening due to hyperfine dipole
interactions, $\Gamma _{en}\sim \rho\mu _e\mu _n$ with 
$\rho=\min\left\{\rho_e,\rho_n\right\}$, the inhomogeneous broadening due 
to nuclear dipole interactions, $\Gamma _{nn}\sim \rho_n\mu _n^2\sim
10^4$ s$^{-1}$, and the parameter $\delta $ is given by Eq. (46), from
which $\delta ^2\sim 10^{-2}\Gamma _{en}^2z_{eff}^2$. The width 
$\Gamma_{en}$, according to Eqs. (69) and (70), is always larger than 
$\Gamma_{nn}$, provided that the density of electrons $\rho_e$ is not
three orders smaller than the density of nuclei $\rho_n$, which follows
from the relation $\Gamma _{en}\sim 10^3\left(\rho_e/\rho_n\right) 
\Gamma _{nn}$. For example, if $\rho_e\sim \rho_n$, then $\Gamma_{en}
\sim 10^3\Gamma _{nn}$. If we take $\rho_e\sim 2\times 10^{20}$ cm$^{-3}$
and $\rho_n\sim \left( 5\times 10^{22}-10^{23}\right) $ cm$^{-3}$, as in
the experiments [6-10], then $\Gamma _{en}\sim (1-10)\Gamma _{nn}$. In
this way, the hyperfine width $\Gamma _{en}$ is usually larger than
$\Gamma _{nn}$ and always larger than $\delta $, although $\Gamma _{en}$
may be comparable with  $\alpha _n$, if $\rho_e\ll\rho_n$. When
$\rho_e\sim \rho_n$, the largest parameters among those considered above
are $\alpha _n$ and $\Gamma _{en}\sim 10^7$ s$^{-1}$.  In such a case,
other parameters entering additively with these can be omitted. For 
instance, the delay time (67) may be written as
$$
t_0=\frac{T_2}{2gs_0}\ln \left| 
\frac{\alpha _n^2+\Gamma _{en}^2}{4\omega _N^2}\right| . 
$$

An important parameter entering the expressions for the characteristic times
and influencing the behavior of solutions is the coupling parameter $g$, 
which is drastically different
for the case when there is magnetic order in the electron system compared
to the case when the magnetization is absent. For the latter case, when $%
z_{eff}=0$, we have $g\sim 10$. When a ferromagnetic material is considered,
so that $z_{eff}\sim 1$, then the second term in Eq. (42) can become much
larger than the first one. Thus, for $\rho_e\sim \rho_n$, we have
$$
\frac{\rho_e\mu _eAz_{eff}}{\rho_n\mu _n\omega _N}\sim \frac{\mu _e}{\mu
_n}\sim
10^3. 
$$

Therefore, the coupling parameter $g$ can be increased by three orders by
the presence of electron magnetization, reaching the value $g\sim 10^4$. The
system of magnetized electrons acts as an additional resonator strongly
strengthening the coupling between the resonance electric circuit and
nuclear spins.

To evaluate the characteristic values of the delay time (67) and relaxation
time (68), we made calculations for several ferromagnetic materials with
typical parameters taken from Ref. [30]. In our formulas we take the
filling factor $\eta =1$, consider the purely resonance case, when $\omega
=\left| \omega _N\right| $, also take $z_{eff}=\frac 12$, $\rho_e=\rho_n$ and
assume that the transverse anisotropy parameter is small compared to 
$\Gamma _{en}$. We analyze the case of purely self-organized coherent
relaxation when at the initial time nuclear spins are polarized against the
external magnetic field, so that $s_0=-I$, where $I$ is an absolute value of
a nuclear spin, and there is no initial coherence imposed upon the system,
so that $u_0=0$. The high initial polarization of nuclear spins can be
achieved by the dynamic nuclear polarization technique. The
transverse relaxation time $T_2=\Gamma _2^{-1}$ can be measured by several
methods, e.g., the two-pulse echo technique or the single-pulse echo
technique [30], of which the former is likely to
be more reliable. Our results are presented in Table 1.

\section{Discussion}

Nonlinear spin dynamics is considered for ferromagnets consisting of electron
and nuclear subsystems coupled through hyperfine forces.
The sample is prepared in a strongly nonequilibrium initial state. In
addition, the ferromagnetic sample is considered inserted into a coil of a
resonant electric circuit. All this makes the spin dynamics highly
nonlinear. The evolution of the system is described by seven nonlinear
equations, six of which are differential equations for electron and
nuclear spins and one equation is an integro-differential equation for the
feedback field of the resonator. These are solved by using the
scale separation approach [11, 12]. It is shown that due to the resonator
feedback field an ultrafast coherent relaxation of spins can occur. The
system of magnetized electrons serves as an additional
resonator for the nuclear spins, significantly enhancing the effective coupling of the nuclear spins with
the resonator circuit. Such an enhancement can reach three orders of
magnitude, as compared to the coupling in a paramagnetic material.

The ultrafast coherent relaxation of nuclear spins may be either triggered
by an initial pulse or can be self-organized. The latter case is the more
interesting, since then all relaxation characteristics, such as the delay
time and relaxation time, depend strongly on the values of the internal
parameters. The most important such parameters, starting the process of
self-organized coherent relaxation and, therefore, defining the main
relaxation characteristics, are the transverse magnetocrystalline anisotropy
and the dipole hyperfine interactions. If the density of electrons is more
than three orders of magnitude lower than the density of nuclei, then the direct nuclear
dipole interactions also become important. An interesting extension of
the present approach could be the inclusion of external alternating magnetic
fields, as has been done for nuclear magnets [33-35]. 

By studying the peculiarities of the coherent spin relaxation, it is possible 
to extract information on the intrinsic properties of magnetic materials. 
This especially concerns the regime of self-organized coherent relaxation, 
whose characteristics are very sensitive to the values of the parameters of 
the material studied. By observing a coherent pulse in the power flow, one 
can measure, with a very high precision, the delay time $t_0$. The latter 
can be accurately measured because it exactly corresponds to the maximum 
of the function $w(t)$, which is proportional to the power flow in the 
resonant circuit [12]. If the delay time $t_0$ is measured 
experimentally, then, inverting (67), one may find the sum of 
$\alpha_n^2$ and $\Gamma_*^2\equiv\Gamma_{en}^2+\Gamma_{nn}^2$ as
$$
\alpha_n^2 +\Gamma_*^2 = 4 \omega_N^2 \exp\left ( 2gs_0\; \frac{t_0}{T_2}
\right ) \; ,
$$ 
where the inequality $\delta\ll\Gamma_{en}$ is taken into account. This 
relation, when $\alpha_n$ 
is known from other experiments, makes it possible to define the 
inhomogeneous broadening $\Gamma_*$. As is mentioned in Sec. VI, one 
usually has $\alpha_n\leq\Gamma_*$. The transverse-anisotropy parameter 
$\alpha_n$ depends on the orientation of the sample with respect to the 
external magnetic field. It is possible to choose an orientation such that
$\alpha_n\ll\Gamma_*$. Then we obtain a simple formula giving 
the inhomogeneous broadening
$$
\Gamma_* = 2\omega_N \exp\left ( gs_0\; \frac{t_0}{T_2}\right )
$$
in terms of the known values of the nuclear magnetic resonance frequency 
$\omega_N$, the coupling parameter $g=g_n$ in Eq. (42), the initial 
nuclear polarization $s_0$, the transverse relaxation time $T_2$, and 
the measured delay time $t_0$. It follows from the analysis of 
Sec. V that $t_0>0$ requires $s_0<0$. Thus the value $s_0t_0$ in the formula 
for $\Gamma_*$ is negative, making $\Gamma_*\ll\omega_N$. The
exponential dependence of $\Gamma_*$ on the delay time $t_0$ makes the 
value of $\Gamma_*$ very sensitive to $t_0$.

Another possibility for exploiting the effect of coherent spin relaxation 
is its sensitivity to initial conditions, in particular, to the initial 
amplitude of the transverse spin, $|u_0|$. The latter, in order to 
influence the delay time (58) and relaxation time (59), should 
be such that
$$
|u_0|^2 > \frac{\alpha_n^2 +\Gamma_*^2}{\omega_N^2}\; s_0^2 \leq 10^{-4} \; .
$$
This is always small, and can be made arbitrarily smaller by reducing $s_0$. 
Hence, we conclude that even quite weak external pulses, resulting in nonzero $|u_0|$, 
can trigger the process of coherent relaxation. For example, from Eqs. (59) and (45) we get
$$
|u_0|^2 =\left ( \frac{T_2}{g\tau_0}\right )^2 +
\left (\frac{\alpha_n^2 +\Gamma_*^2}{\omega_N^2} - 1\right ) \; s_0^2 \; .
$$
This allows us, by measuring the coherent relaxation time $\tau_0$, to 
find the initial amplitude $|u_0|$. The sensitivity of coherent spin 
relaxation to initial conditions could be employed for creating ultrasensitive
detectors of weak external pulses. In turn, this can be used to construct 
sensitive particle detectors [19].

\vskip 3mm

In conclusion, the main results obtained in this paper can be summarized as:

\vskip 2mm

(i) A theory of nonlinear spin dynamics is developed for the systems of 
electron and nuclear spins coupled with each other through hyperfine 
forces and also coupled to a resonator electric circuit. 
This essentially generalizes the previous consideration of 
nonequilibrium nuclear magnets [11--13] to a much wider class of 
materials, having long-range magnetic order.

\vskip 2mm

(ii) The very complicated set of nonlinear differential equations 
is solved by invoking the scale separation approach [11, 12]. It 
is important that, because of the existence of small parameters resulting 
in different time scales, the motion of electron and nuclear spins can be 
effectively separated, as is seen in Eqs. (47)--(50).

\vskip 2mm

(iii) The effect of a strong {\it coupling-parameter enhancement}, due to 
the presence of magnetic order in the electronic subsystem, is described. 
The effective nuclear coupling parameter can be 
enhanced by three orders of magnitude, which makes relaxation really ultrafast, with 
the relaxation time (63) becoming smaller than $T_2$ by four orders of magnitude.

\vskip 2mm

(iv) The nature of all main intrinsic mechanisms triggering the self-organized
coherent relaxation, that occurs in the absence of external pulses, is 
elucidated. These mechanisms, defining the delay time (67), are the 
electron-nuclear interactions through hyperfine dipole forces, nuclear 
dipole interactions, and magnetocrystalline anisotropy fields.

\vskip 2mm

(v) Two types of applications are discussed.
One type concerns the investigation of the internal properties of the materials 
by measuring the delay time (58) and coherent relaxation time (59). 
Another type utilizes the sensitivity of these 
characteristic times to the initial conditions, giving the  
possibility of employing coherent spin relaxation for 
the ultrasensitive detection of weak external pulses.

\vspace{5mm}

{\bf Acknowledgments}

\vspace{2mm}

One of the authors (V.I.Y.) is grateful for useful discussions to N.A.
Bazhanov. Financial support from the University of Western Ontario and
NSERC of Canada is appreciated.

\vspace{1cm}

{\Large{\bf Appendix. Feedback Field}}

\vspace{5mm}

According to the scale separation approach [11, 12], we first consider 
Eq. (19) for the fastest variable $x$, treating there the slow
variables $s$ and $z$ as quasi-integrals of motion. Because of the second
inequality in Eq. (32), we can, in a first approximation, omit the term
containing $H$, whereupon the solution of (19) is
$$
x\simeq \left( x_0-\frac{A_ez}{\Omega _e+i\gamma _2}\right) \exp \left\{
i\left( \Omega _e+i\gamma _2\right) t\right\} + 
\frac{A_ez}{\Omega_e+i\gamma_2}. 
$$
with $\Omega _e\equiv \omega _e-As-\xi _0$ and $A_e\equiv \alpha _e-Au-\xi$. In 
Eq. (22) we keep, as a quasi-integral of motion, the slow variable $s$.
Because of the second inequality in Eq. (32), we again, to the first
approximation, may omit the term with $H$. Averaging, in the right-hand side
of Eq. (22), the electron variables over the period $2\pi /\omega $, we get an
approximate equation
$$
\frac{du}{dt}\simeq i\left( \Omega _n+i\Gamma _2\right) u-iA_ns, 
$$
where $\Omega _n\equiv \omega _n-Az_{eff}-\varphi _0$ and $A_n\equiv \alpha_n-Ax_{eff}-\varphi$. We find
$$
u\simeq \left( u_0-\frac{A_ns}{\Omega _n+i\Gamma _2}\right) \exp 
\left\{i\left( \Omega _n+i\Gamma _2\right) t\right\} +
\frac{A_ns}{\Omega _n+i\Gamma_2}. 
$$
Substituting the approximate expressions into Eq. (25), we come to the
form
$$
H=-2{\rm Re} \left[ \beta _e\left( t\right) \frac{dx}{dt}+
\beta _n\left( t\right) \frac{du}{dt}\right] , 
$$
in which
$$
\beta _e\left( t\right) =\pi \eta \frac{\mu _e\rho_e}{\delta _e}\left [
\exp \left ( \delta _et\right) -1\right) , \qquad
\beta _n\left( t\right) =\pi \eta \frac{\mu _n\rho_n}{\delta _n}\left [
\exp\left ( \delta _nt\right) -1\right ] , 
$$
$$
\delta _e=i\left( \omega -\omega _e+As+\xi _0\right) +\gamma _2-\gamma _3, 
\qquad \delta _n=i\left( \omega -\omega _n + Az +\varphi _0\right) +
\Gamma_2-\gamma_3. 
$$
The functions $\beta _e\left( t\right) $ and $\beta _n\left( t\right) $ do
not vary much during the period $2\pi /\omega $, so we may replace them by their averages over this time,
$$
\beta _e\equiv \frac \omega {2\pi }\int\limits_0^{2\pi /\omega }\beta
_e\left( t\right) dt=\pi ^2\eta \frac{\mu _e\rho_e}\omega \left(
1+\frac{2\pi
\delta _e}{3\omega }\right) , 
$$
$$
\beta _n\equiv \frac \omega {2\pi }\int\limits_0^{2\pi /\omega }\beta
_n\left( t\right) dt=\pi ^2\eta \frac{\mu _n\rho_n}\omega \left(
1+\frac{2\pi
\delta _n}{3\omega }\right) . 
$$
Omitting here the small terms $\delta _e/\omega $ and $\delta _n/\omega $,
we obtain Eq. (36).

\newpage

{\bf Table 1}. The characteristic parameters related to the self--organized
coherent nuclear spin relaxation in several ferromagnetic materials.

\vspace{1cm}

{\begin{table} \centering
\begin{tabular}{|l|l|l|l|l|l|l|l} \hline
{Sample}  &  {Nucleus}  &  {$I$}  &  {$\hspace{0.2in}\omega_N$}  &  {\hspace{0.2in}$T_2 $} &  {\hspace{0.2in}$\tau_0$} &  {\hspace{0.2in}$t_0$} \\ 
$$ & $$ & $$ &  $(10^9$ Hz) &  $(10^{-4}$ s) & $(10^{-8}$ s) & $(10^{-8}$ s) \\ \hline
Li$_{0.5}$Fe$_{2.5}$O$_4$ & $^{57}$Fe & $1/2$ & $0.47$ &  $40$ &$88.2$ &$400$  \\
Mn$_{0.51}$Sb$_{0.49}$O$_4$ &  $^{123}$Sb & $7/2$ & $1.31$ & $1.70$ &$0.54$ &  $2.98$  \\
$$ & $^{55}$Mn  & $5/2$  &  $1.61$ &  $0.60$ &  $0.26$ &  $1.53$ \\
NiMnSb & $^{55}$Mn &  $5/2$ &  $1.88$ &  $0.95$ &  $0.42$ &  $2.49$\\
NiMnSi & $^{55}$Mn &  $5/2$ &  $2.01$ &  $0.60$ &  $0.26$ &$1.59$ \\
Co$_2$MnSi &  $^{59}$Co &  $7/2$  &  $0.91$ &  $0.38$ & $0.12$ & $0.62$ \\
$$  & $^{55}$Mn  &  $5/2$  &  $1.59$ &  $0.80$ &  $0.35$ &  $2.03$ \\
Co (fcc) & $^{59}$Co &  $7/2$  &  $1.37$ &  $0.30$ &  $0.09$ & $0.53$ \\
Co (hcp) & $^{59}$Co &  $7/2$  &  $1.38$ &  $0.65$ &  $0.21$  & $1.15$\\ \hline
\end{tabular}

\end{table}}

\end{document}